\documentclass[9pt, a4paper, twocolumn]{article}
\usepackage[a4paper, total={7.5in, 10.5in}]{geometry}
\usepackage{xurl}

\usepackage[super,comma,sort&compress]{natbib}
\usepackage{abstract}

\usepackage{lipsum}
\usepackage{amsmath,amssymb}
\usepackage{graphicx}
\usepackage{verbatim}
\usepackage{caption}
\usepackage{subcaption}
\usepackage[noblocks]{authblk}
\usepackage[table,xcdraw]{xcolor}
\usepackage{braket}
\usepackage{titlesec}
\usepackage{float}

\titleformat{\section}
  {\normalfont\fontsize{12}{15}\bfseries}{\thesection}{1em}{}

\usepackage{hyperref}
\hypersetup{colorlinks=true, urlcolor=blue, linkcolor=blue, citecolor=blue}

\title{Clifford and Non-Clifford Splitting in Quantum Circuits: Applications and ZX-Calculus Detection Procedure}
\author{Fernando Lima}
\author{Arcesio Castañeda Medina}

\affil[1]{Fraunhofer ITWM}

\begin{document}


\twocolumn[
  \begin{@twocolumnfalse}
    \maketitle
    \begin{abstract}
      \noindent Classical simulation of quantum circuits is a pivotal part of the quantum computing landscape, specially within the NISQ era, where the constraints imposed by available hardware are unavoidable. The Gottesman-Knill theorem further motivates this argument by accentuating the importance of Clifford circuits and their role on this topic of simulation. In this work, we propose and analyze use cases that come from quantum circuits that can be written as product between a Clifford and a Non-Clifford unitary, these ranging from fully classical emulation, hybrid quantum-classical execution or even quantum algorithm simplification. To further complement this analysis, we make use of ZX-Calculus and its assets to detect a limiting border of these circuits that would allow for a separation between a Clifford section and a Non-Clifford section. To achieve this, we present a novel procedure for parsing ZX diagrams, that not only allows for the detection of this border but also simplifies the circuit extraction process. 
      \newline
      \newline
    \end{abstract}
  \end{@twocolumnfalse}
]


\section{Introduction}

Quantum computing within the NISQ era is limited by various constraints that result from the limitations of the machinery used to implement it. These range from qubit fidelities\cite{Sanders_2016}, algorithm scalability\cite{beverland2022assessingrequirementsscalepractical, chen2022complexitynisq}, noise\cite{Wang_2021, Mih_likov__2022}, and several other issues whose improvements are reliant on the evolution of available technologies. For this reason, even at the cost of potential "quantum advantage", classical simulation of quantum circuits is of utmost importance in this field, either it be for validation of quantum hardware, or for testing and verification of quantum algorithms\cite{xu2023herculeantaskclassicalsimulation}.

In particular, the case of stabilizer circuits is of great interest, as these represent a class of circuits that can be efficiently simulated on classical hardware\cite{Aaronson_2004}, meaning that quantum advantage is not lost in this case. Quantum error correction\cite{gottesman1997stabilizercodesquantumerror} and magic state distillation\cite{Bravyi_2005} are some of the most common uses for this class of circuits, but its use on accelerating classical simulation via piece-wise simulation of more general circuits containing Clifford fragments\cite{xu2023herculeantaskclassicalsimulation} is also a great indicator on how the properties of these kinds of circuits might contribute to a quantum/classical splitting of more general circuits, which is what partially motivates this work.

In this work, we present a procedure for parsing ZX diagrams which simultaneously identifies a cutoff border that separates a Clifford section of the circuit starting on its inputs or outputs, and a Non-Clifford section, and we further discuss on how we can leverage this procedure for quantum/classical splitting via several use cases that come from having our circuit in a $U_CU_{NC}$ or a $U_{NC}U_C$ format, with $U_C$ corresponding the unitary representing the Clifford Section, and $U_{NC}$  representing the Non-Clifford one. Sections \ref{ZX-Calculus}, \ref{Classical Simulation and Stabilizer Circuits} serve as the theoretical ground for this work, with the former containing general information on ZX-Calculus and its relevant properties and procedures that are used throughout this work, while the latter is a brief discourse on stabilizer circuits. Section \ref{Clifford Border Detection Algorithm} contains the description of the Clifford Border Detection procedure and sections \ref{Circuits in U_NCU_C format} and \ref{Circuits in U_CU_NC format} and their subsections contain all the results and discussion regarding potential applications of having the circuit split into a $U_CU_{NC}$ or a $U_{NC}U_C$ unitary sequence respectively. The topics and analysis of these scenarios range from classically simulatable stabilizer states as well as their initialization and properties, in sections \ref{Stabilizer Canonical Form}, \ref{Generator Projector Initialization} and \ref{Projector Circuit Distribution}, classical statevector simulation in section \ref{Stabilizer Statevector for Classical Simulation} and VQE applications as well as estimation of expectation values in sections \ref{Split Clifford+T Ansatz Fed VQE} and \ref{Expectation Value Estimation}. Finally, section \ref{Conclusions and Future Work} serves as a summary of this work, with a brief reiteration of the main results and some reflection on possible future work.


\subsection{ZX-Calculus}
\label{ZX-Calculus}

ZX-Calculus is a graphical language commonly used in several topics of the quantum computing realm (MBQC, circuit optimization, validation, etc.), mainly due to its practicality in representing and manipulating linear maps, and hence corresponding quantum circuits.

The underlying agents of this representation are called ZX-Diagrams, which are composed by a set of generators connected by wires and whose shape is malleable based on a set of rules.
The most common type of generators are called spiders and they come in two types.

Z-spiders are n-input and m-input green objects, equipped with a scalar $\alpha$, representing a linear map of the form:

$$
\ket{0}^{\otimes m}\bra{0}^{\otimes n} + e^{i\alpha}\ket{1}^{\otimes m}\bra{1}^{\otimes n}
$$

While X-spiders are n-input and m-input red objects, equally equipped with a scalar $\alpha$, representing a linear map of the form:

$$
\ket{+}^{\otimes m}\bra{+}^{\otimes n} + e^{i\alpha}\ket{-}^{\otimes m}\bra{-}^{\otimes n}
$$

The graphical depiction of these generators is, respectively:

\begin{figure}[H]
    \centering
    \includegraphics[scale=1]{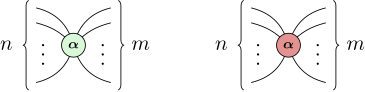}
    \caption*{}
    \label{Spiders}
\end{figure}

Additionally, due to its prevalence, there is a special representation for the linear map corresponding to the Hadamard Gate, instead of the sequence of spiders that would otherwise represent it:

\begin{figure}[H]
    \centering
    \includegraphics[scale=0.35]{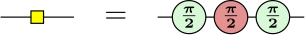}
    \caption*{}
    \label{Spiders}
\end{figure}

A more concise representation of the Hadamard generator is commonly used, and will henceforth be used in this work as well, in which it is represented by a dashed blue wire referred to as an Hadamard wire, depicted as follows:

\begin{figure}[H]
    \centering
    \includegraphics[scale=1]{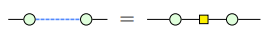}
    \caption*{}
    \label{Spiders}
\end{figure}

With these generators, it is possible to represent any $2^n \times 2^m$ complex matrix from some $n$ and $m$, and hence, any linear map between qubits\cite{vandewetering2020zxcalculusworkingquantumcomputer}. This is known as the universality if ZX-Calculus.

While being able to use these ZX-Diagrams to represent linear maps that correspond to circuits and states is very useful in itself, the true power of ZX-Calculus lies in the ability to manipulate these diagrams via a set of rewrite rules that allow for derivation of different diagrams while preserving the linear map that is represented. There are several sets of rules that may be used for this purpose, with a common set\cite{Kissinger_2020} being the one we will use throughout this work which is shown in figure \ref{Spiders}.

\begin{figure}[H]
    \centering
    \includegraphics[scale=0.74]{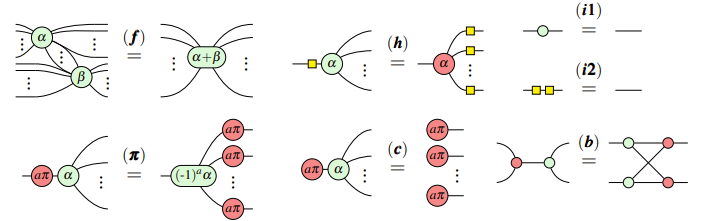}
    \caption{Set of ZX-Calculus rewrite rules}
    \label{Spiders}
\end{figure}

One particular set of diagrams that is greatly important to this work is the set of Graph-like ZX-Diagrams, whose form will allow for application of a circuit extraction procedure that will be introduced shortly. In accordance to its definition\cite{Duncan_2020}, a ZX-Diagram is graph-like when:

\begin{itemize}
    \item It only contains Z-spiders
    \item Its Z-spiders are exclusively connected via Hadamard wires
    \item It has no parallel Hadamard wires or self-loops
    \item All of its inputs and outputs are connected to a Z-spider and all of its Z-spiders are at most connected to one input or output
\end{itemize}

Given a ZX-Diagram in graph-like form, it is possible to perform an extraction procedure to write the diagram in circuit-like form, meaning that it can easily be translated into circuit form. We will give a brief, step-by-step overview of this procedure, but we refer to Backens et al.\cite{Backens_2021} for a full in-depth description.

\begin{enumerate}
    \item Define the frontier of the diagram, this being a set of green spiders such that everything to its right looks like a circuit. This means that the initial frontier will be the rightmost set of spiders, as nothing in the diagram will yet be circuit-like, but as the algorithm is being applied, this frontier gets updated and a circuit will be built on its right.
    \item Unfuse any connection between frontier spiders as a CZ gates
    \item Extract any possible Clifford operator present on the frontier spiders (phases and Hadamards)
    \item Define the biadjacency matrix of the bipartite graph consisting of frontier spiders on the right, and the rest of the unextracted graph on the left. Proceed with a full Gaussian elimination on this matrix, adding a CNOT between wires $a$ and $b$ of the extracted circuit for each sum modulo 2 operation between rows $a$ and $b$ during this procedure.
    \item Once there are no spiders to extract besides the frontier spiders, either we go back to step 2 for the next iteration, or if there is no more spiders to extract to the left of the frontier, the procedure terminates.
\end{enumerate}

As it is clear to see, at the end of the procedure we should have a ZX-Diagram containing Hadamard edges and spiders, and each spider (or pair of spiders in the case of vertically connected ones) translates directly into an unitary gate, and hence the diagram can be represented as a circuit.


\subsection{Classical Simulation and Stabilizer Formalism}
\label{Classical Simulation and Stabilizer Circuits}

Classical simulation of quantum circuits is not only a prevalent subject in the realm of quantum computing, but it is actually the preferred approach in collecting information of quantum state generated by certain classes of quantum circuits. One such class would be the class of stabilizer circuits, which correspond to circuits that can be perfectly simulated in in polynomial time by a probabilistic classical computer, according to the well-known Gottesman-Knill Theorem\cite{gottesman1998heisenbergrepresentationquantumcomputers}. These circuits are defined by the following properties:

\begin{enumerate}
    \item The circuit contains Clifford group gates only
    \item The qubits of the circuit must be prepared in computational basis states
    \item Any measurement in the circuit must be done in the computational basis
\end{enumerate}

Essentially, given a n-qubit circuit initialized to state $\ket{\psi_{i}} = \ket{0}^{\otimes n}$ and containing only a set of gates that generates the Clifford group, such as the set of CNOT, Hadamard and Phase gates, it is possible to simulate this circuit classically via efficient procedures such as Clifford Tableaus\cite{Aaronson_2004, Gidney_2021}, granted that measurements are done in the computational basis.

The states that result from these stabilizer circuits are known as stabilizer states and are of great importance, particularly in the field of quantum error correction\cite{gottesman1997stabilizercodesquantumerror}. Stabilizer states in an $2^n$-dimensional Hilbert Space are special states that can be uniquely identified by their stabilizer group, this being a $2^n$-dimensional set of $n$-dimensional linearly independent Pauli strings which stabilize the state, i.e. from which these states are eigenvectors with eigenvalues +1.

While the size of the stabilizer group of a given $2^n$-dimensional state is $2^n$, any $n$-dimensional presentation of this group is enough to uniquely identify the state, as each linearly independent Pauli string consecutively bisects the Hilbert space in equal halves, one being generated by the correspondent +1 eigenvalued eigenvectors, and the other by the -1 eigenvalued eigenvectors. This means that if we keep track of the subspace generated by the positive eigenvalues, after $n$ splits, we are left with a 1-dimensional subspace which necessarily describes the stabilizer state\cite{qubitguide}. Table \ref{8DHilbertBisection} contains a representation of these successive splittings via linearly independent Pauli strings, also called stabilizer generators, for an 8-Dimensional Hilbert space (3-qubits). This can be used to visualize the several subspaces that are created at each bisection, and also the subspace represented by the stabilizer state, highlighted in green.

\begin{table}[h]
\centering
\begin{tabular}{|
>{\columncolor[HTML]{C0C0C0}}l |
>{\columncolor[HTML]{9AFF99}}l lllllll|}
\hline
{\color[HTML]{333333} $G_0$} & + & + & + & + & - & - & - & - \\
{\color[HTML]{333333} $G_1$} & + & + & - & - & + & + & - & - \\
{\color[HTML]{333333} $G_2$} & + & - & + & - & + & - & + & - \\ \hline
\end{tabular}
\captionsetup{font=footnotesize}
\caption{Pictorial representation of the bisections of the $2^3$-dimensional Hilbert space produced by the linearly-independent generators $G_0$, $G_1$ and $G_2$. The column in green represents the common subspace that is stabilized by all the generators, which identifies the stabilizer state of said generators}
\label{8DHilbertBisection}
\end{table}

It is worth mentioning that when using the Clifford Tableau method to efficiently simulate a stabilizer circuit, no information regarding the resultant statevector is attained, with the output of this algorithm being a set of stabilizer generators that stabilize this state.

In practice, these successive partitions of the Hilbert space can be quite useful, particularly if we wish to project a general quantum state into the common stabilized subspace created by these splits. The key to this lies in the fact that, given a set of n  stabilizer generators $\{G_1,...,G_n\}$, the state $\ket{\psi_{st}}$ that is stabilized by such set can be reproduced via a projection of a random state $\ket{\psi_r}$ onto the +1 eigenspace of the generators\cite{Aaronson_2004} (noting that this random state must not be perpendicular to the resultant stabilizer state), which can be achieved by the following operation:

\begin{equation}
    \ket{\psi_{st}} = \frac{1}{2^n}\prod_{i=1}^n(I + G_i)\ket{\psi_r}
    \label{StabProj}
\end{equation}

While this allows us to theoretically recreate the superposition that is stabilized by a given generator set, the main agent behind this operation, this being the $I + G_n$ operator, is not unitary, and so it cannot be reproduced as a set of gates on a quantum circuit.

This problem can be addressed via different approaches, one such being block encoding of unitaries\cite{9951292}, but a more straightforward way is to use a projector circuit, a common practice in quantum error correction\cite{gottesman2009introductionquantumerrorcorrection} 

\begin{figure}[H]
    \centering
    \includegraphics[scale=0.8]{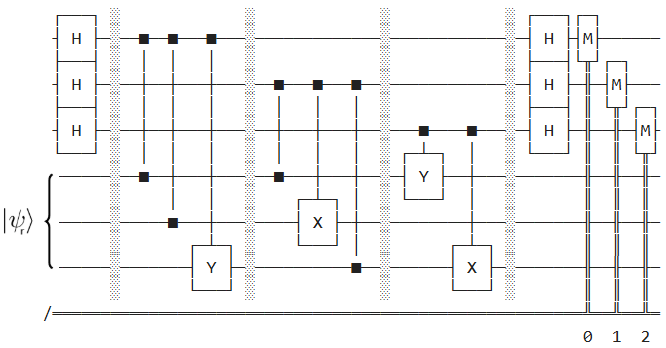}
    \caption{Example of post-selection generator projector circuit for a a 3-qubit initial state $\ket{\psi_r}$ and the stabilizer generator set of \{+ZZY, +ZXZ, +YIX\}. The barriers in the circuit delimit each controlled version of a generator.}
    \label{8DProjCirc}
\end{figure}

With our target qubits initialized in the state $\ket{\psi_r}$, we would need then an ancilla qubit per generator. Focusing solely on one generator $G$ and its respective ancilla qubit, we would start from the following state:

$$
\ket{\psi} = \ket{0}\otimes\ket{\psi_r} 
$$

This would be followed by an Hadamard gate on our ancilla qubit, which would create the following state:

$$
\ket{\psi} = \frac{1}{\sqrt{2}}(\ket{0}+\ket{1})\otimes\ket{\psi_r} 
$$

A controlled version of the generator $G$ would then be applied through the qubits, such that:

$$
\ket{\psi} = \frac{1}{\sqrt{2}}(\ket{0}\otimes\ket{\psi_r} + \ket{1}\otimes G\ket{\psi_r})
$$

And a final Hadamard gate on the ancilla qubit would give:

\begin{align*}
 \ket{\psi} & = \frac{1}{2}\big[\ket{0}\otimes\left(\ket{\psi_r} + G\ket{\psi_r}\right) + \ket{1}\otimes \left(\ket{\psi_r} - G\ket{\psi_r}\right)\big]\\
 & = \frac{1}{2}\big[\ket{0}\otimes\left(I + G\right)\ket{\psi_r}+ \ket{1}\otimes\left(I - G\right)\ket{\psi_r}\big]
\end{align*}

And hence, we can follow up with a measurement of the ancilla qubit, which will either result in a collapse to $\ket{0}$ or $\ket{1}$, with the former ensuring that the remaining qubits will be in the state $\left(I + G\right)\ket{\psi_r}$, meaning that our initial state was projected on the positive eigenspace of G, and the latter will be in the state $\left(I - G\right)\ket{\psi_r}$, and hence a projection onto the negative eigenspace of G occurred.

If we then consider the entire stabilizer set, we could stack these projector circuits sequentially granted we add an ancilla qubit per generator, and hence would be able to project our initial state onto the common positive eigenspace of all generators. This would happen when the measurement of our n ancillas results in $\ket{0}^{\otimes n}$ and our post measurement state of the target qubits would be the stabilizer state $\ket{\psi_{st}}$.

Figure \ref{8DProjCirc} depicts an example of such a circuit for a 3-qubit initial state $\ket{\psi_r}$, and with the target stabilizer generator set being \{+ZZY, +ZXZ, +YIX\}. If we consider these to represent the generators $G_0$, $G_1$ and $G_2$ present in table \ref{8DHilbertBisection}, it becomes clear to see that each measurement result of the ancilla qubits refers to one of the subspaces generated by the bisections represented in this table, with a measurement of $\ket{000}$ corresponding to the projection onto the stabilizer subspace higlighted in green.

\section{Clifford Border Detection Algorithm}
\label{Clifford Border Detection Algorithm}

In this section, we describe the procedure we developed in order to identify the Clifford Border of a circuit-like ZX-diagram resulting from the extraction procedure described in section \ref{ZX-Calculus}.

Starting from an extracted circuit-like diagram, we will have a general Clifford+T ZX-diagram containing both green and red spiders, the latter ones coming from the CNOT's introduced by the extraction procedure. Our first intent will be to try and push any Non-Clifford spider, i.e. any spider whose phase is not an integer multiple of $\frac{\pi}{2}$, as far right as possible on the diagram. One naive approach to do so, and the one used in this work, is to convert any red spider on the diagram to green making use of the $\textbf{(h)}$ rule of figure \ref{Spiders}, and then iteratively use the spider fusion and unfusion rule, depicted as the $\textbf{(f)}$ rule of figure \ref{Spiders}, on the Non-Clifford spiders when possible, commuting them with adjacent green spiders until it is no longer possible.

The next step consists on identifying the fragment of each wire that separates the Clifford and Non-Clifford sections of the circuit. To this end, we start by collecting a list of vertices that were parsed while our Non-Clifford pushing was being done, such that every vertex that is located on the same wire of our Non-Clifford spiders and is to the left of it after the pushing is saved.

At this moment, we already have a preliminary splitting of circuit, such that our soft Clifford Border is composed precisely by the last vertex of each wire that is present in our list of parsed vertices, which will either be a Non-Clifford spider or an output spider. This splitting, however, is not ideal, as we are neglecting the fact that there could be two-qubit gates which connect a wire that is on the Clifford section to a wire on the Non-Clifford section. To account for this, we consider any such gate to be entirely on the Non-Clifford section, updating the list of parsed vertices and the border definition accordingly. Nevertheless, the update on the list of parsed vertices may create additional occurrences of new two-qubit gates being in-between the two sections. To this end, we must recursively apply this updating procedure until no such situation occurs.

\begin{figure}[]
    \centering
    \includegraphics[scale=0.5]{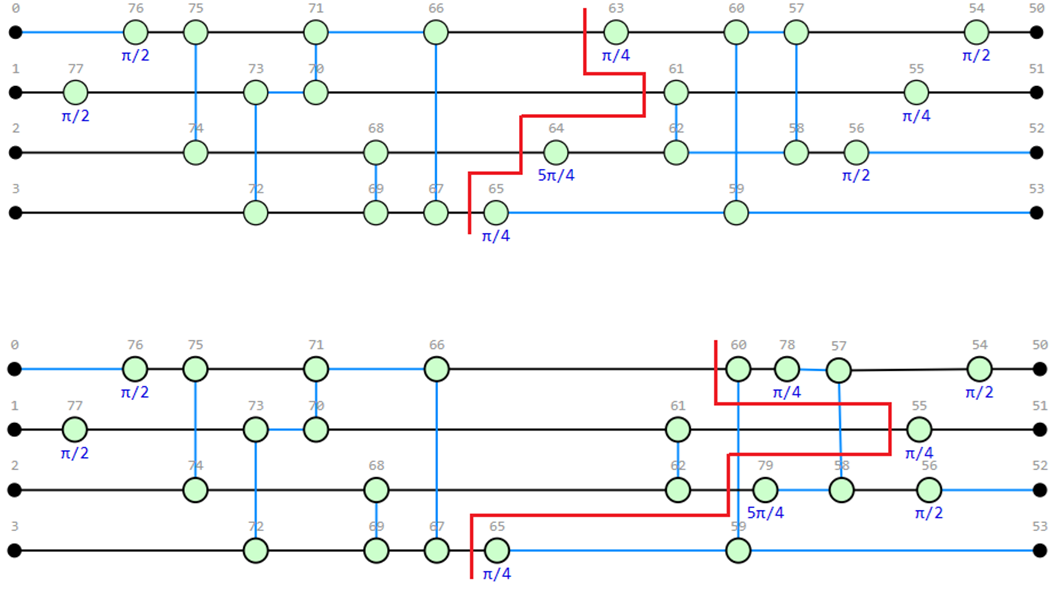}
    \caption{Example of the Clifford Border on a 4-qubit Clifford+T circuit before (up) the pushing procedure and after (down) the pushing procedure}
    \label{CBorder}
\end{figure}

Once this is done, a clear border that separates the Clifford and Non-Clifford sections of the circuit is defined. One such example, applied to a 4-qubit Clifford+T circuit, can be seen in figure \ref{CBorder}. This figure also highlights the importance of the pushing procedure in order to minimize our Non-Clifford section, as it is clear to see that certain gates that were originally on the Non-Clifford section might actually be moved to the Clifford section, as is the case with the CZ gate composed by vertices 61 and 62 of the diagram.

As it stands, this procedure is relatively simple and obviously quite dependent on the shape of the circuit, as the placement of the Clifford border is determined by the position of Non-Clifford gates along the circuit, which makes it so that any circuits containing Non-Clifford gates close to the initial input qubits will probably generate a negligible Clifford section. It is then direct to assume that, without further operations on top, this procedure would be best suited for smaller circuits containing few Non-Clifford operations. Nonetheless, properties of the unitaries composing the circuits might tie to scenarios in which we could further maximize the Clifford section before proceeding with the border identification, and hence represent potential improvements to this procedure. One such example lies in the so-called quantum simulation circuits\cite{liu2025quclearcliffordextractionabsorption}, commonly used as ansatzes in VQE and QAOA optimization problems, which contain Non-Clifford subsections given by exponentiated Pauli Strings that weakly commute with Cliffords\cite{debrugière2024fastershortersynthesishamiltonian}.

\section{Results and Discussion}
\label{Results and Discussion}

The focal point of this work, as stated previously, is to go over different use cases associated with circuits which can be split into a Clifford and a Non-Clifford section, represented by a $U_C$ unitary and a $U_{NC}$ unitary respectively. The procedure described in the previous section was developed with the intent of enabling this splitting, and it serves as a naive method to clearly identify both unitaries, regardless if the circuit is written in a $U_{C}U_{NC}$ format, or a $U_{NC}U_{C}$ format. Both of these formats have distinct characteristics and applications which we will go over in the following sections, but one common metric among these use cases is the relative size of the Clifford circuit. 

With this in mind, we ran our border detection procedure on several benchmark circuits with the intent of checking the relative size of their left and right clifford sections. The set of circuits chosen for this is composed of most of Qiskit's QASMBench\cite{10.1145/3550488} small circuits, and most of 2 to 5 qubits circuits from MQT Bench\cite{Quetschlich_2023}.

\begin{figure}[]
    \centering
    \includegraphics[scale=0.7]{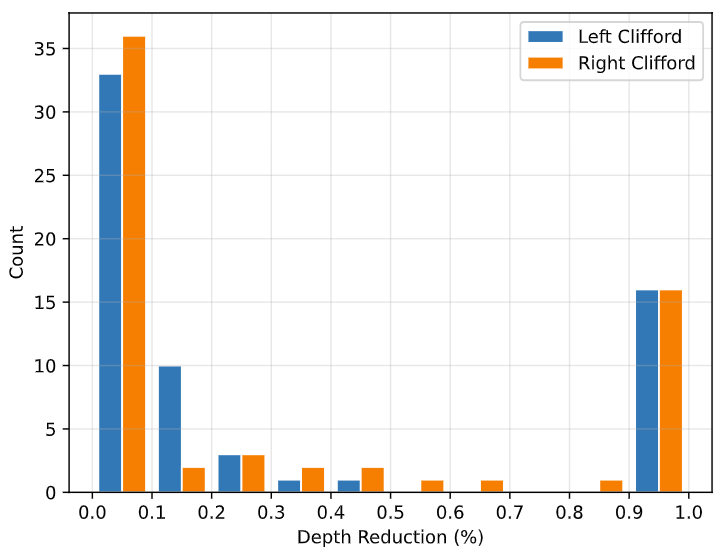}
    \caption{Counts for each percentile corresponding to the relative depth reduction of Qiskit's QASMBench small circuits and MQT Bench 2 to 5 qubit circuits}
    \label{DepthReductionBench}
\end{figure}

\begin{figure}[]
    \centering
    \includegraphics[scale=0.7]{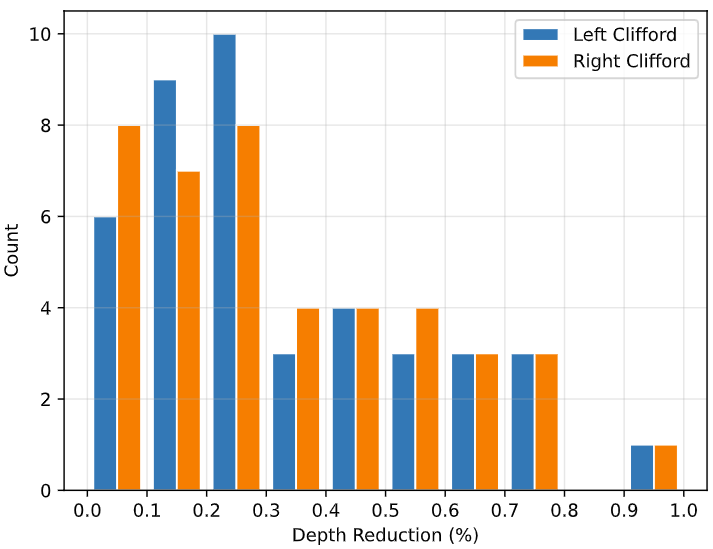}
    \caption{Counts for each percentile corresponding to the relative depth reduction of randomly generated Clifford+T circuits with 2 to 10 qubits, 20 to 100 initial depth and T-Gate probability of $20\%$}
    \label{DepthReductionRand}
\end{figure}

Figure \ref{DepthReductionBench} shows how the depth of the circuit to run on quantum hardware varies after application of the procedure. It is clear to see that most counts lie in the $0\%$ to $10\%$ range, and a significant number is also represented on the $90\%$ to $100\%$ range. The latter case refers mostly to fully Clifford circuits, and hence we disregards talking about them as we are focusing on circuits containing both Clifford and Non-Clifford sections. As for the former, it refers to circuits whose Clifford section is small or non-existent, with a big percentage of these being typical Quantum Simulation circuits used as ansatzes in optimization problems, and hence their Clifford section might be susceptible to an increase if, as mentioned before, an appropriate procedure is taken into account before the border detection\cite{liu2025quclearcliffordextractionabsorption}. The remaining cases refer to all of the scenarios in between, with an obvious decrease in count of cases as the percentage of depth reduction increases. While these cases represent a smaller relative number of circuits compared to the two aforementioned extreme cases, it is important to consider that these are the result of running the version of our procedure described in section \ref{Clifford Border Detection Algorithm}, without any additional optimization steps that could potentially relate to an increase in the Clifford section, and yet, they still indicate the existence of situations in which this splitting can be leveraged in order to attain better time and/or accuracy efficiency via a number of use cases, as we will describe in further sections.

Nonetheless, since the number of circuits present in this category is relatively small, using them as a means of benchmarking some of the use cases is less reliable. Since we are mostly focusing on the relative depth of the Clifford section, we then make use of randomly generated Clifford+T circuits with both number of qubits and depths being similar to the ones present in the benchmark circuits portrayed in Figure \ref{DepthReductionBench}, these ranging from 2 to 10, and from 20 to 100 respectively, and with a T-Gate probability of $20\%$.

Figure \ref{DepthReductionRand} shows then how the relative depth reduction behaves for this class of circuits, and it is clear that the distribution is quite different than the one present in Figure \ref{DepthReductionBench}. However, the pattern of the distribution depicting a decrease in counts with an increase in depth reduction is still present, and this class will allow us to have more accurate benchmarks in the next sections for the circuits that lie in the $10\%$ to $80\%$ depth reduction range.


It is of course important to take into account that, in order for the depth reduction to be relevant, the existence of an efficient protocol to replace the Clifford circuit is required. In the next sections, we will go over on how this issue might be addressed for varying cases stemming from the splitting that either originates a $U_{NC}U_C$ circuit, or a $U_CU_{NC}$ one.

\subsection{Circuits in $U_{NC}U_C$ format}
\label{Circuits in U_NCU_C format}

We start by addressing the case where the Clifford section of our circuit is to the left, which corresponds to the default scenario presented in our splitting procedure described in section \ref{Clifford Border Detection Algorithm}. As the Clifford section is applied to the ground state, we are in accordance with the constraints defined by the Gottesman-Knill theorem, and hence the main concern regarding the use cases for this kind of circuits will be to find an appropriate initialization protocol for the stabilizer state generated by the Clifford section, such that the Non-Clifford section can be fed that state.

\subsubsection{Stabilizer Canonical Form}
\label{Stabilizer Canonical Form}

One very relevant result in what concerns stabilizer circuits would be the gate complexity of stabilizer circuits in canonical form\cite{Aaronson_2004}, which states that any unitary stabilizer circuit has an equivalent "canonical" circuit with only $O(n^2/\log{n})$ gates.

\begin{figure}[]
    \centering
    \includegraphics[scale=0.6]{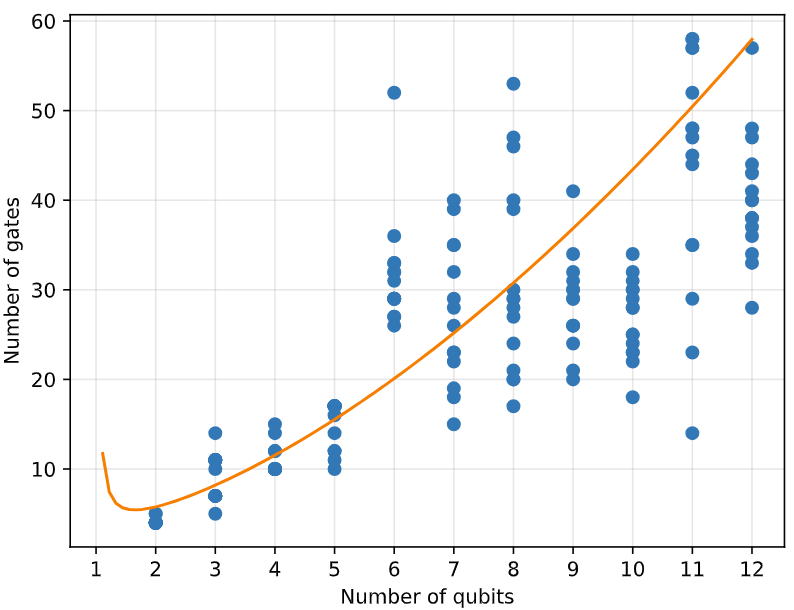}
    \caption{Number of gates of each randomly generated Clifford+T circuit with 2 to 12 qubits and 20 to 400 initial depth versus the $n^2/\log{n}$ canonical threshold}
    \label{CanonForm}
\end{figure}

While it is difficult to discuss how these canonical circuits look like and exactly how many gates they should have, we can still take a look at their scaling curve and compare it with the number of gates of the stabilizer circuits resultant from our splitting procedure, as shown in figure \ref{CanonForm}.

From this figure, we can extract that most of the resultant stabilizer circuits are at a configuration that respects the number of gates of the canonical circuit corollary, as they are placed below the limiting curve. Nonetheless, the existence of circuits whose number of gates stands above this curve might be an indication that there are cases in which the stabilizer circuit is not yet in this form, and hence this form could serve as the initialization protocol that would reduce the gate count and depth of the original circuit.


\subsubsection{Generator Projector Initialization}
\label{Generator Projector Initialization}

As noted in section \ref{Classical Simulation and Stabilizer Circuits}, it is possible to implement the expression in equation \ref{StabProj} by building an appropriate projector circuit that outputs a post measurement state correspondent to a stabilizer state, when given the correspondent set of generators. Indeed we can acquire these generators by running the Clifford Tableau of the extracted Clifford part, making it possible for us to build this circuit

While the number of qubits of the new circuit will be double the original amount, as we need an ancilla qubit per generator, the depth of this new circuit before transpilation is a fixed interval, dependent on the number of identity gates on the generators, ranging from 3 (2 Hadamard layer + Measurement layer) and 3 + $n^2$, as there are $n$ generators, each with a maximum of $n$ non-identity controlled operations.

Given this, any situation in which the depth of a stabilizer circuit sits above the upper limit of depth for these generator circuits corresponds to a case in which it could be advantageous to initialize the stabilizer state via this circuit. It is, however, crucial to verify if this is a common occurrence, as the stabilizer section of larger circuit is, most of the times, a relatively small portion of the original circuit, as we've discussed at the start of this section.

 \begin{figure}[]
     \centering
     \includegraphics[scale=0.7]{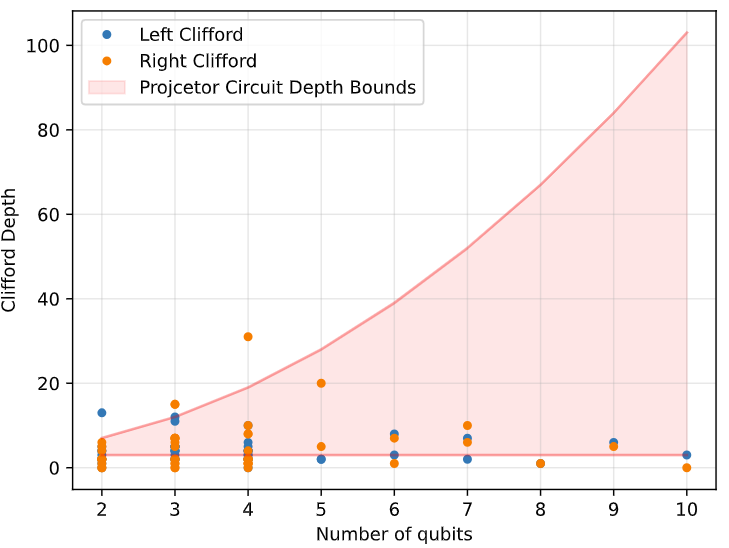}
     \caption{Depth of the Clifford section for Qiskit's QASMBench small circuits and MQT Bench 2 to 5 qubit circuits, as well as the limiting area regarding the depth of potential projector circuits for given number of qubits}
     \label{DepthDiff}
 \end{figure}

 Figure \ref{DepthDiff} shows the depth of the extracted Clifford section of the previously mentioned circuit from QASMBench and MQT Bench, with blue points indicating the Clifford extracted from the left and the orange points indicating the Clifford extracted from right. The red area corresponds to the area where the depth of the projector circuit will lie, being bounded by the aforementioned minimum boundary of 3 and maximum boundary of 3 + $n^2$.

 The minimum boundary of 3 is a fringe case in which the Pauli Strings of the generators are entirely composed of identity gates, and hence represents an unrealistic scenario, while the upper bound corresponds to a case in which none of these strings contain an identity. Hence, for the circuits located inside the boundaries, it is plausible to say that the closer they are to the upper boundary, the more likely they are to be in a scenario in which the projector circuit has a smaller depth than that of the original Clifford. Additionally, any points sitting above the upper boundary correspond to situations in which the projector circuit is guaranteed to be smaller, making them the best candidates for replacement when using this metric. It is clear that the number of cases in which this occurs is small, and it seems to become a more unlikely scenario as the number of qubits grows, as we can only identify matches up to 4 qubits. Nonetheless, we reinforce the idea that these numbers come from the direct splitting of the circuits without further optimizations, and hence the possibility for improvements exists and is likely.
 


 It is worth noting that while we are using the depth as a general metric to analyze this comparison, there are several other factors that could still invalidate the generator circuit as a valid initialization procedure, such as the fact that it requires several ancilla qubits, the amount of two qubit gates and most notably, the necessity of correcting the post measurement state when the measurement of the ancilla qubits is not $\ket{0}^{\otimes n}$.

\subsubsection{Stabilizer Statevector for Classical Simulation}
\label{Stabilizer Statevector for Classical Simulation}

Another relevant instance in which this Clifford splitting could be useful is the case of full classical statevector simulation. In this context, we argue that proceeding with the splitting of the stabilizer section and its simulation separate from the non-Clifford section would be faster than running a statevector simulation of the entire circuit. The fundamental point behind this reasoning is the fact that methods such as Clifford Tableaus allow for a significant speed up when simulating stabilizer circuits, compared to a statevector simulation, but unfortunately, it would then become unfeasible to simulate the non-Clifford section of the original circuit, as the result of our tableau is not itself a statevector that we can evolve from, but instead a group of stabilizer generators.

One accurate way to reproduce the statevector produced by the stabilizer circuit via its generators would be through the operation described in equation \ref{StabProj}, but as mentioned before, it is impossible to write a circuit that performs such operation due to the non-unitary nature of it, and so, any attempt to reproduce the statevector using an alternative circuit falls under the discussion of the previous two sections.

It is possible, however, to classically compute a desired n-qubit stabilizer state by using an algorithm that runs a randomly generated n-qubit statevector through the $n$ projector operations depicted in equation \ref{StabProj}. Such algorithm has the benefit of being independent of the circuit responsible for generating the stabilizer state, as it only concerns itself with the generators, and hence its complexity only varies with the number of qubits.

A precise example of an algorithm that does as explained above would be Stim's\cite{Gidney_2021} implementation of their \textit{to\_state\_vector} function, which can be called by a \textit{Tableau} object in order to convert the latter into an array description of a statevector. With this, we have the necessary tools to acquire the stabilizer statevector generated by the stabilizer section of our circuit post splitting, and evolve it through the Non-Clifford section in order to produce the output statevector of the full circuit.

\begin{figure}[]
    \centering
    \includegraphics[scale=0.7]{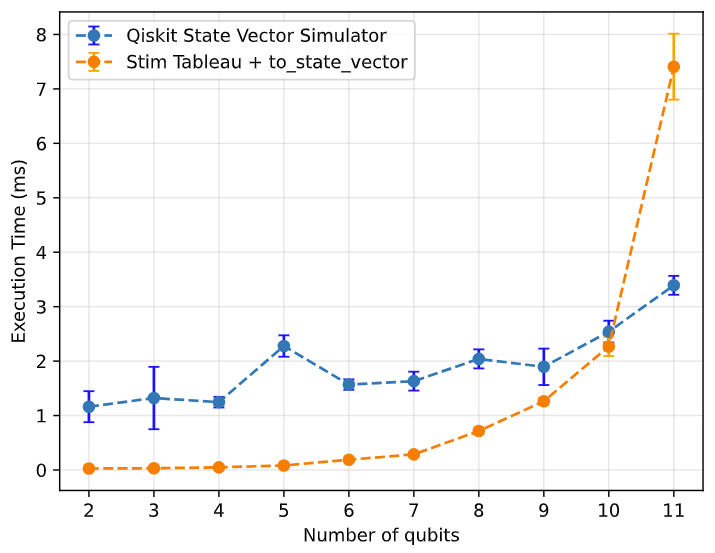}
    \caption{Execution time of Qiskit's Statevector simulator for randomly generated stabilizer circuits with qubit numbers between 2 and 11, and with 40 depth, against the execution time of Stim's implementation of the Clifford Tableau followed by transformation to statevector array via \textit{to\_state\_vector} function.}
    \label{StimSV_vs_QiskitSV}
\end{figure}

\begin{figure}[]
    \centering
    \includegraphics[scale=0.73]{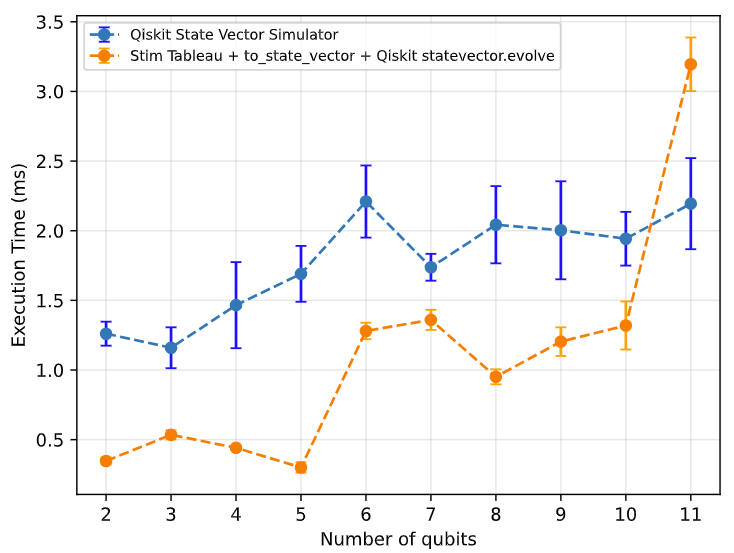}
    \caption{Execution time of Qiskit's Statevector simulator for randomly generated Clifford+T circuits with qubit numbers between 2 and 11, T-Gate probability of 20\% and with 40 depth, against the execution time of same simulator for Non-Clifford section of the circuits initialized by stabilizer statevector obtained via Stim's implementation of the Clifford Tableau followed by transformation to statevector array via \textit{to\_state\_vector} function.}
    \label{CircSim_Stim_v_Qiskit}
\end{figure}

Figures \ref{StimSV_vs_QiskitSV} and \ref{CircSim_Stim_v_Qiskit} illustrate the usefulness of using this algorithm in the context of statevector simulation. Figure \ref{StimSV_vs_QiskitSV} depicts the comparison between execution times of Qiskit's AerSimulator implementation of an exact statevector simulator and the execution times of Stim's implementation of the Clifford Tableau followed by transformation to a statevector array via the \textit{to\_state\_vector} function. The circuits used for these simulations were randomly generated stabilizer circuit with a fixed depth of 40 and with number of qubits varying from 2 to 11, and the plotted points and error bars respectively correspond to the average and variance of a set of 7 runs for each number of qubits and with 50 loops per run. The output statevector array of both procedures is, as expected, the same.

While it is expected that the simulation of a stabilizer circuit via Clifford Tableau is faster than the simulation of the same circuit via a statevector simulator, it might not be evident that the statevector description of the output state can be obtained quicker via the former, and indeed that is not the case in every situation, as we can clearly see from the exponential growth in execution time due to increase in number of qubits depicted in the figure, with the orange plot points representing a larger average execution time relative to the blue plot points starting at the mark of 11 qubits. This can be explained due to the $\Omega (n^22^n)$ complexity of the algorithm used by the \textit{to\_state\_vector} function\cite{desilva2024fastalgorithmsclassicalspecifications}, which reveals the sensitivity of such procedure to larger numbers of qubits, as expected due to the fact that not only the number of generators, and hence projectors, is the same as the number of qubits, but also the number of superposition elements of our initial random state, onto which these generators have to be applied, increases by a factor of two per qubit. Given this exclusive dependence on the number of qubits, which sets both the number of generators and the number of state superposition elements, in contrast to a dependence of circuit size and shape, we should note that based on our runs, the speedup behavior seems to be prevalent regardless of depth, and hence while we only show here the results regarding circuits with 40 depth, similar results are expected when varying this quantity. Nonetheless, it is clear that if the execution time is the decisive metric, using the tableau and projector conversion to the statevector is the preferred method when it comes to statevector simulation of stabilizer circuits up until 10 qubits.

Given these results and the preliminary discussion at the beginning of this section, it becomes of immediate interest to try and leverage this speedup for a more general class of circuits by using the splitting procedure introduced in the previous sections. Figure \ref{CircSim_Stim_v_Qiskit} depicts the comparison between execution times of, once again, Qiskit's AerSimulator statevector implementation, and the execution times of a composite circuit simulation which consists of:

\begin{enumerate}
    \item Application of Stim's Clifford Tableau to the stabilizer section of the circuit, obtained through our splitting procedure, followed by transformation to a statevector array via the \textit{to\_state\_vector} function.
    \item Transformation of the statevector array into a Qiskit Statevector object, and processing of this object via the \textit{Statevector.evolve} method, with the Non-Clifford section of the circuit as the operator to evolve by.
\end{enumerate}

Once again, the circuits used for these simulations were randomly generated Clifford+T circuits with a fixed depth of 40, number of qubits varying from 2 to 11, and T-Gate probability of 20\%, and the plotting scheme follows the same logic as that of Figure \ref{StimSV_vs_QiskitSV}.

\begin{figure}[]
    \centering
    \includegraphics[scale=0.73]{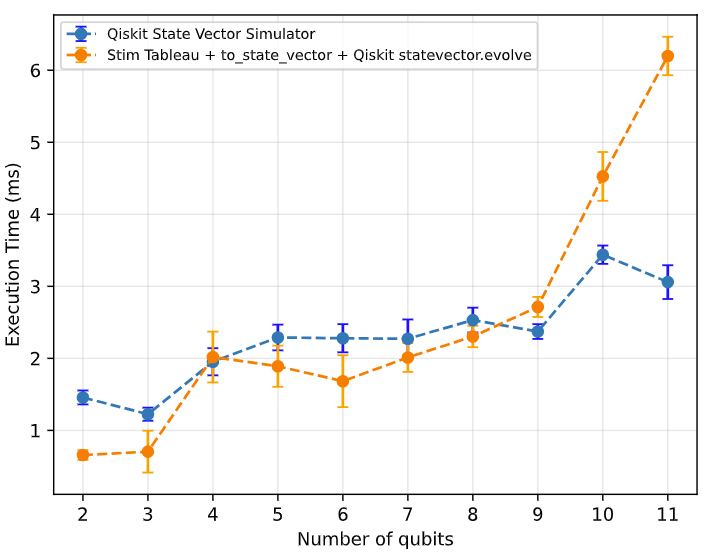}
    \caption{Execution time of Qiskit's Statevector simulator for randomly generated Clifford+T circuits with qubit numbers between 2 and 11, T-Gate probability of 20\% and with 100 depth, against the execution time of same simulator for Non-Clifford section of the circuits initialized by stabilizer statevector obtained via Stim's implementation of the Clifford Tableau followed by transformation to statevector array via \textit{to\_state\_vector} function.}
    \label{CircSim_Stim_v_Qiskit_100}
\end{figure}

In alignment with the results outlined in Figure \ref{StimSV_vs_QiskitSV}, the clearest result depicted by this figure is the reduction in execution time corresponding the composite circuit simulation mentioned above for a number of qubits between 2 and 10, with a turn occurring at around the 11 qubit mark. The considerations regarding the comparison between the two sets of data are then very similar to the ones already discussed, as it is clear that as the number of qubits grows, the execution time of the procedure for the generation of the initialization statevector starts to take over as the main contributor for the overall time, and hence, even if the section of the circuit that effectively runs on the statevector simulator is smaller when the splitting is done, the overall time of running all the parts will be larger than just running the full circuit on the statevector simulator,as can be seen for a qubit count of 11.

Another important feature to discuss regarding these results is their behavior for different circuit depths. As opposed to the results shown in Figure \ref{StimSV_vs_QiskitSV}, both methods make use of Qiskit's statevector simulator, with the first method using it fully, and the second only using it for a section of the circuit, which is relevant in what concerns the slight behavioral variations that we see represented as local minima and maxima spikes along the orange plot points. Since the purpose of our method is to reduce the Non-Clifford section to be simulated on the statevector simulator, the time advantage that we identify along the plot points is precisely due to the fact that generating an initialization statevector for the stabilizer state does indeed save time, but since the circuits we are using are randomly generated, the depth of the stabilizer section of these circuits can widely vary, as we have shown in Figures \ref{DepthReductionBench} and \ref{DepthReductionRand}. Because of this variance, we can expect that the time saved from running our composite simulation method will be proportional to depth of the stabilizer section resulting from the splitting, which explains the spiky behavior that we observe on our data, as certain plot points might be susceptible to the fact that a larger stabilizer section was extracted, and hence the time saving is larger, resulting on a local minima, while on the other hand, some points might correspond to circuits in which the stabilizer section was smaller, resulting in local maxima. This effect becomes largely important when discussing how this difference in execution times scales with depth, as in general, when the depth of our Clifford+T circuits increases, the percentage of circuit corresponding to a stabilizer section is smaller, which in turn makes it so that the section that is to be run on a statevector simulator is asymptotically similar to the entire unsplit circuit, and hence the execution times of both methods will be significantly closer up until the point in which the initialization statevector generation process takes over. This point is illustrated in Figure \ref{CircSim_Stim_v_Qiskit_100}, in which the execution times are obtained in the same conditions as those present in Figure \ref{CircSim_Stim_v_Qiskit}, but with the depth of the circuits now set to 100.

Regardless of this discussion, it is evident in both Figures \ref{CircSim_Stim_v_Qiskit} and \ref{CircSim_Stim_v_Qiskit_100} that up until a number of qubits, our proposed composite procedure is sistematically better for statevector simulation than Qiskit's current implementation of regular statevector simulation in what concerns time efficiency for a select class of circuits.

\subsubsection{Split Clifford+T Ansatz Fed VQE}
\label{Split Clifford+T Ansatz Fed VQE}

Given the advantages identified in the previous section when it comes to statevector simulation, it becomes immediate to wonder about possible use cases in which this speedup may be leveraged.

Out of the several ways one can choose in order to run a VQE algorithm, using statevector simulation for posterior expectation value computation is a common sight, and so, with the appropriate conditions in place, we now expect that, by adapting a standard implementation of such algorithm to take into account our circuit splitting procedure as well as our proposed statevector initialization, an improvement concerning time efficiency arises.

Indeed if we take a look into the formalism of a general VQE algorithm, given an operator represented observable $A$, and an ansatz $\ket{\psi(\Vec{\theta})} := U(\Vec{\theta})\ket{\psi}$ where $U(\Vec{\theta})$ is a parameterized circuit and $\ket{\psi}$ is a given initial state, the overall purpose would be to compute the ground state eigenstate of our observable by employing a variational approach on the parameters of our ansatz, whose evolution is dictated according to the minimization of a cost function $f(\Vec{\theta})$ given by:

\begin{equation}
    f(\Vec{\theta}) = \bra{\psi(\Vec{\theta})}A\ket{\psi(\Vec{\theta})}
    \label{lossfunction}
\end{equation}

Typical optimization procedures, whose function is to iteratively compute sets of parameters that minimize this function, usually employ the method of gradient descent, which makes use of the local rate of change of the cost function in parameter space on each iteration, such that the next set of parameters can be determined. The update rule for the $n^{th}$ iteration of such procedure, with a scalar value $s$ as a step size parameter, is commonly given by:

\begin{equation}
    \Vec{\theta}_{n+1} = \Vec{\theta}_{n} - s\Vec{\nabla}f(\Vec{\theta}_{n})
    \label{NextStep}
\end{equation}

In order to showcase the potential gains of using our splitting procedure in the context of VQE, we will assume that our ansatz circuits are Clifford+T circuits, in which the T-Gates are replaced by parameterized Z-rotation gates. Our splitting will then make it so that we can separate a purely Clifford section $U_{C}$, this being our typical stabilizer circuit, and a Non-Clifford parameterized section $U_{NC}(\Vec{\theta})$, such that:

\begin{equation}
    U(\Vec{\theta}) = U_{NC}(\Vec{\theta})U_{C}
\end{equation}

With our ansatz circuits written in this form, it becomes clear on how our procedure can be used in order to speedup operations that will be executed along the optimization protocol. For once, if we consider our n-qubit initial state $\ket{\psi}$ as $\ket{0}^{\otimes n}$ (which we will abbreviate as only $\ket{0}$ for simplicity), we can rewrite our loss function $f(\Vec{\theta})$ shown in \ref{lossfunction} as:

\begin{equation}
  \begin{split}
    f(\Vec{\theta}) & = \bra{0}U^{\dag}_{C}U^{\dag}_{NC}(\Vec{\theta})AU_{NC}(\Vec{\theta})U_{C}\ket{0}\\
    & = \bra{\psi_{st}}U^{\dag}_{NC}(\Vec{\theta})AU_{NC}(\Vec{\theta})\ket{\psi_{st}}
    \end{split}  
    \label{SplitLF}
\end{equation}

Where $\ket{\psi_{st}} := U_{C}\ket{0}$ is the stabilizer state produced by the Clifford section $U_{C}$ of the original circuit, and hence, can be quickly and efficiently simulated via the methods discussed in the previous sections.

Similarly for the gradient element of Equation \ref{NextStep}, it is easy to run through some computations and achieve a similar expression yielding:

\begin{equation}
    \Vec{\nabla}f(\Vec{\theta}) = \bra{\psi_{st}}\Vec{\nabla}[U^{\dag}_{NC}(\Vec{\theta})AU_{NC}(\Vec{\theta})]\ket{\psi_{st}}
    \label{SplitLFgrad}
\end{equation}

Seeing as both Equations \ref{SplitLF} and \ref{SplitLFgrad} represent operations that will be executed mulitple times along a VQE run, both requiring simulation of the ansatz circuit, it becomes relevant to discuss the possibility of speeding up the process by computing the stabilizer state $\ket{\psi_{st}}$ a priori and using it to speedup the simulation, allowing for an expected reduction in execution time of the procedure.

We will be testing this starting from Qiskit's native implementation of the VQE algorithm (via their VQE class), using the Hamiltonian for the $H_2$ molecule as our operator (using Qiskit Nature's implementation of this Hamiltonian constituted by a set of Pauli operators), from whom we intend to compute its ground state. The ansatz circuits will then be randomly generated 2-qubit Clifford+T circuits with varying depths.

Since we intend to include the method discussed in the previous section as an alternative for the simulations done along the run, a few of modifications in the implementation of some of Qiskit's scripts were done:

\begin{enumerate}
    \item The VQE class was changed so that it may receive a Statevector object as an initialization argument, which in turn will be passed along the chain of functions that are called in order to perform the necessary computations.
    \item The Simulator instance initialized inside the Estimator was changed so that it only runs when no initialization stabilizer statevector is passed. If one is indeed passed, then the simulation is replaced by the composite method presented in the previous section.
    \item The native implementation for the computation of expectation values is built so that it works for any kind of simulation method, but since we are focusing only on statevector simulation (which is the default simulation method of the Estimator class), we instead compute the expectation values directly from the statevector resultant from the simulation, making use of the \textit{expectation\_value} function of the \textit{Statevector} class, which is considerably faster.
\end{enumerate}

\begin{figure}[]
    \centering
    \includegraphics[scale=0.35]{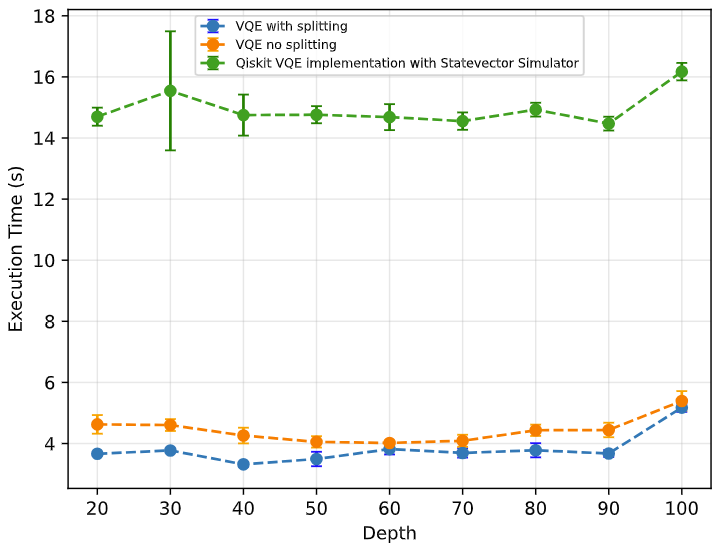}
    \caption{Execution times of VQE procedure for the $H_2$ molecule using randomly generated Clifford+T ansatzes with varying depths and with T-Gate probability of 20\%. The green plot points correspond to the native Qiskit implementation of the procedure, the orange and blue plot points correspond to a modified implementation using \textit{Statevector.expectation\_value} with the latter receiving the initialization stabilizer statevector as an argument}
    \label{VQE_ExeTimes}
\end{figure}

One additional aspect to consider when running this VQE is the dependence of its performance and accuracy on the depth of the ansatz circuit, since there are several factors that go into the selection of this value. Firstly, as we discussed in the previous section, the time speedup that we get from our alternative statevector simulation is sensitive to the depth of the circuit, and so we also expect this effect to be present here, but secondly, it is relevant to consider several different depths in order to account for the fact that the number of parameters in the ansatz circuit will very likely be proportional to its depth, as we are randomly generating circuits with a fixed probability of such parameters. Given that the accuracy of the algorithm is based on the selection of the set of these parameters that can best approximate the ground state of our Hamiltonian, it is expected that a larger number of parameters allows for a larger set of output states, and hence the probability to get a better approximation is higher.

Figure \ref{VQE_ExeTimes} depicts the execution of the full run of the VQE procedure for the $H_2$ molecule. As mentioned above, the ansatz circuits are randomly generated Clifford+T circuits, with depths varying from 20 and 100 and with T-Gate probability of 20\%. Each set of plot points, distinguishable by their colors, refers to a different configuration, with the green plot points corresponding to a run using Qiskit's native implementation of the procedure with the simulator set to the default method, that being statevector, and with the orange and blue plot points corresponding to the modified implementation using the \textit{Statevector.expectation\_value} of the \textit{Statevector} class, with the latter receiving the initialization stabilizer statevector as an argument, and hence making use of the alternative statevector simulation described in the previous section. As is the case in the figures presented in previous section, the plotted points and error bars respectively correspond to the average and variance of a set of a number of runs per unit, with several loops each. In this case, these correspond to set of 7 runs for unit of depth and with 5 loops per run.

It is clear to see that, while not an effect of our splitting procedure, the use of the \textit{Statevector.expectation\_value} function is the main factor in what concerns speeding up the execution of the algorithm relative to its native implementation on Qiskit, which in itself is a relevant result that, to our knowledge, has not been discussed in other sources, and could represent a direct improvement on the current performance of the procedure for the particular circumstance of statevector simulation, given that the precision seems to be preserved as we will discuss further.

As for the other two sets, the trend seems to be quite similar to the one represented in figures \ref{CircSim_Stim_v_Qiskit} and \ref{CircSim_Stim_v_Qiskit_100}, as even the time difference between both sets seems to be proportional to the difference noted in those figures, with the timescales differing due to the fact that multiple circuit runs occur over the course of the procedure. The effect of decreasing time savings as the depth grows seems to also be present, alas at a smaller scale, but even this seems to be in accordance with the results of last section, as the circuits being used are 2-qubit circuits, and out of all the experiments ran for the benchmarking done in that section, these are the least impacted by that aspect. All in all, the general idea conveyed by these results is, as expected, the same as the one acquired from the last section, which indicates that the usage of the splitting and subsequent processing of the circuit elements according to our proposition seems to be, for time efficiency purposes, best suited for this scenario than the current method.

Regarding the accuracy of the several runs, as the main purpose of these results is not to evaluate the quality of the ansatzes used, but to discuss about the advantages provided by our method when using them, we mention only that both the least and the most accurate scenarios for the two sets conveying the modified expectation values (which have the same outputs, as their simulation only differs due to the initialization), and the set conveying the native implementation, correspond to the same depths of 30 and 100 respectively, with the first two methods presenting a deviation from the reference eigenstate value of 0.62397 for the former, and 0.02031 for the latter, while the native method presents deviations of 0.62060 and 0.01783 respectively. While both these values present slightly smaller deviations concerning the native method, this occurrence is not fixed as several deviation values regarding other depths have the opposite behavior, and different sets of runs with different circuits can also result in the opposite trend. What seems to hold regardless of the selected circuits and their depths is the precision between results, as can be noticed by the similarity in the deviation values presented. In fact, out of all the results per depth of the various runs, the largest difference between the output values of the two modified methods and the native method is of 0.00904, which for reference, represents only 0.48\% of the tabulated value.


\subsection{Circuits in $U_CU_{NC}$ format}
\label{Circuits in U_CU_NC format}

We now look into the case where the Clifford section of the circuit is instead located on the right. In order to achieve this with our proposed detection procedure, we would simply need to adjoint our target circuit and run the procedure on the adjoint circuit. After the detection of the border, we would then adjoint the circuit back and split at the border. In this scenario, the Clifford section would not be in accordance with the restraints of the Gottesman-Knill theorem as the state in which it would act would not be a stabilizer state, but as we will see with the use cases, it is possible to still make use of this theorem in order to simplify the execution.

\subsubsection{Expectation Value Estimation}
\label{Expectation Value Estimation}

One of the biggest use cases for this scenario can be identified by taking into account the VQE description presented in Section \ref{Split Clifford+T Ansatz Fed VQE} with the order of the unitaries switched. The in-depth description of this use case with possible applications on Quantum Simulation Circuits is already thoroughly tried\cite{liu2025quclearcliffordextractionabsorption}, and hence we will simply go through a brief explanation of this idea.

Starting from the loss function defined in Equation \ref{lossfunction}, we now write our ansatz as $U(\Vec{\theta}) = U_{C}U_{NC}(\Vec{\theta})$, such that the function to minimize can be written, in terms of the observable $A$, as:

\begin{equation}
  \begin{split}
    f(\Vec{\theta}) & = \bra{0}U^{\dag}_{NC}(\Vec{\theta})U^{\dag}_{C}AU_{C}U_{NC}(\Vec{\theta})\ket{0}\\
    & = \bra{0}U^{\dag}_{NC}(\Vec{\theta})A'U_{NC}(\Vec{\theta})\ket{0}\\
    \end{split}  
    \label{SplitRF}
\end{equation}

With $A' = U^{\dag}_{C}AU_{C}$. From this expression, we determine that the function given by the expectation value of $A$ on the full circuit is the same as the expectation value of the Clifford transformed observable $A'$ on the Non-Clifford circuit. Noting that any Hermitian observable can be written as a sum of Pauli observables, and that the n-dimensional Clifford group normalize the n-dimensional Pauli group\cite{mastel2023cliffordtheorynqubitclifford}, then the Paulis that constitute the sum of our general observable A will remain in the Pauli group after transformation through $U_C$. Equation \ref{SplitRF} then suggests that the Clifford section can be entirely removed from the estimation of the expectation value, as long as the measurement basis is corrected according to the transformation applied to our observable (or the Paulis that constitute it).

This then translates into a considerable reduction of the workload to be sent to the quantum backend responsible for computing these values, at the cost of requiring to compute the update of the Pauli observables to be measured, which in itself represents a quite efficient classical procedure with $O(n^2)$ complexity\cite{liu2025quclearcliffordextractionabsorption}.

\subsubsection{Projector Circuit Distribution}
\label{Projector Circuit Distribution}

While using a generator projector circuit is suitable for replication of a stabilizer state by replacing the original Clifford, as described in section \ref{Generator Projector Initialization}, we propose a different way of using these circuits that also serve as replacement of the Clifford section in case we want to replicate the probability distribution generated by a circuit in the $U_CU_{NC}$ format.

To understand how this can be achieved, we start by noting that our Non-Clifford circuit acting on the initial state $\ket{0}^{\otimes n}$ can be written as a superposition of the elements of a given orthonormal basis in the $2^n$-dimensional Hilbert space. As such, an interesting choice of basis would be $U_c^{\dag}\ket{n}$, where $\ket{n}$ represents an element of the computational basis. It is clear to see that this transformation yields a valid orthonormal basis, as the identity $\bra{m}U_CU_C^{\dag}\ket{n} = \delta_{nm}$ is valid. Given that, we have:

\begin{align*}
    U_CU_{NC}\ket{0} & = U_C\sum_na_nU_c^{\dag}\ket{n}\\
    & = \sum_na_n\ket{n}
\end{align*}

The expression above suggests that the weights given to our selected basis before executing the Clifford section are the same weights given to the computational basis after executing the Clifford. It is also clear that the probability distribution generated from running the circuit will be given by the squared absolute value of these weights $|a_n|^2$.

We proceed by noticing that we can run the sum generated by the $U_{NC}$ unitary as the following:

\begin{align*}
    \sum_na_nU_c^{\dag}\ket{n} &= \sum_na_nU_c^{\dag}X_n\ket{0}\\
    & = \sum_na_nX'_nU_c^{\dag}\ket{0}\\
    & = \sum_na_n\ket{\psi_{st}^n}
\end{align*}

Where $X'_n = U_c^{\dag}X_nU_c$ and $X_n$ is  the n-dimensional Pauli string made up of the tensor product between $X$ and $I$ gates such that $X_n\ket{0} = \ket{n}$. Noting that n-dimensional Cliffords normalise the n-dimensional Pauli group\cite{mastel2023cliffordtheorynqubitclifford}, we have that $X'_n$ is also a Pauli String.

We now focus on the stabilizer generator set of $\ket{\psi_{st}^0} = U_c^{\dag}\ket{0}$, which as discussed previously, corresponds to a set $S = \{G_1, ..., G_n\}$ of linearly independent Pauli Strings that simultaneously stabilize the state. The effect of these generators on the other basis elements $\ket{\psi_{st}^n}$ of the superposition will be as follows:

\begin{align*}
    G_i\ket{\psi_{st}^n} & = G_iX'_nU_c^{\dag}\ket{0}\\
    & = \begin{cases} 
      +\ket{\psi_{st}^n} &  [G_i, X'_n] = 0\\
      -\ket{\psi_{st}^n} &  \{G_i, X'_n\} = 0
   \end{cases}
\end{align*}

Where we use the fact that any pair of n-dimensional Pauli strings either commute or anti-commute\cite{qubitguide}.

Knowing that our $\ket{\psi_{st}^n}$ are orthogonal, table \ref{8DHilbertBisection} gives us a good example on how generators split a n-dimensional space in terms of n positive and negative eigenvalued orthogonal subspaces, each represented by a + or - string of n elements, meaning that our sum in terms of $\ket{\psi_{st}^n}$ represents a superposition of states in which each lives in one of the subspaces generated by the generators of $\ket{\psi_{st}^0}$. It is also relatively simple to determine this + and - string due to the fact that the generators of $\ket{\psi_{st}^n}$ are nothing more than the generators of $\ket{0}^{\otimes n}$ transformed by $Uc^\dag$, which would originally be the group generated by the n independent strings ZI...I, IZ...I, ..., II...Z. Since the same transformation is applied to both this group and the $X_n$, we need only to compute the original commutation relations between each $X_n$ with all $G_n$ in the computational basis, as they will be preserved. Doing this sequentially through the $G_i$, updating our string with a + for each commutation and a - for each anti-commutation, yields the string indicating the generated subspace, which will coincidentally be directly related to the shape of our $X_n$ with I and X corresponding to + and - respectively.

What this means is that our state is then, as mentioned above, a superposition of the states distributed among the subspaces created by the bisections of the Hilbert space by the generators $S = \{G_1, ..., G_n\}$, with each element weighted by $a_n$ and the corresponding subspace being given by the string determined by $X_n$, i.e. $II...I \xrightarrow{}++...+$, $XI...I \xrightarrow{}-+...+$, $IX...I \xrightarrow{}+-...+$ and so on.

If we now look back at the projector circuits mentioned in section \ref{Classical Simulation and Stabilizer Circuits}, using the Non-Clifford state as the initial state and the generators $S = \{G_1, ..., G_n\}$, the measurement of the ancilla qubits will result in a post measurement state that will fall in one of these subspaces created by the generators, with the result of the measurement corresponding to the + and - string associated with the space, 0 and 1 corresponding to + and - respectively. Since these states are weighted by the $a_n$, with each of these weighing the corresponding $n^{th}$ state which in turn is mapped to the measurement of $\ket{n}$ in the ancilla qubits, we get that the distribution produced by the measurement of the ancilla qubits of this circuit is the same as the distribution generated by the original circuit $U_CU_{NC}\ket{0}$.

This results makes it so that we have a similar procedure of replacing the Clifford section of our circuit with a possibly advantageous projector circuit, just like in section \ref{Generator Projector Initialization}, with the added benefit that there is no need to correct the post-measurement state, as here we are looking for the distribution generated by the ancilla qubits. Nonetheless, the reasoning regarding the pros and cons of using such replacement circuit need to be taken into account and we refer to discussion presented in the aforementioned section, as the replacement and its features are analogous.

\section{Conclusions and Future Work}
\label{Conclusions and Future Work}

The primary focus of this work was to showcase the potential advantages of leveraging a circuit written as a sequence of a Clifford unitary and a Non-Clifford unitary. We then started by introducing a Clifford Border detection procedure and put forward some use cases in which it can be used, as it could seemingly represent a reduction of circuits that are to be executed on quantum machines, or even simulated on classical ones. As we have shown, Qiskit's QASMBench and MQT Bench small circuits contain several scenarios in which a significant uninterrupted section of the circuit corresponds to a Clifford unitary, which may compose a relative depth value ranging from $10\%$ up to scenarios of $90\%$, as can be seen in Figure \ref{DepthReductionBench}. These are the results of running our border detection procedure that works by identifying the initial section of the circuit that is exclusively built by Clifford gates and which has no entanglement with any section that contains Non-Clifford operations, and so it is worth noting that the relative depth value has room for improvement by possibly introducing commutation rules that would allow to maximize the Clifford section before the splitting.
While our procedure focuses only on finding a singular split, by isolating the biggest possible Clifford subcircuit starting from the inputs or outputs, another possible improvement to explore would be the identification of further Clifford subsections. The major difficulty with this improvement would be to argue on how we could leverage these subsections in order to improve our simulation metrics, with some possible procedures on this being mentioned along the paper, such as the weakly commuting property of certain Non-Clifford circuits with Cliffords\cite{liu2025quclearcliffordextractionabsorption}, but other possible alternatives being the existence of generalizations beyond stabilizer circuits that are efficiently simulatable and even simulations of stabilizer circuits with non-stabilizer initial states\cite{Aaronson_2004}.

From our analysis of our border detection procedure on the aforementioned benchmark circuits and also randomly generated Clifford+T circuits with varying depths, we were able to argue how useful this procedure can be in certain occasions, as even complex initialization protocols such as the ones conveyed by equation \ref{StabProj} could be adapted to potentially simpler quantum circuits whose execution results in the stabilizer state we want to reproduce in order to feed the Non-Clifford section of the circuit. It is indeed important to note that the kind of circuits we examined in section \ref{Generator Projector Initialization} quickly become unfeasible as the number of qubits grows, and so our argument regading these circuits only holds for a smaller set of circuits whose qubit number is within the 2 to 5 range, as it is clear in figure \ref{DepthDiff}. Nonetheless, a very intriguing subject for improvement would be to explore alternatives on how to implement the operation described by equation \ref{StabProj} as a quantum circuit, ideally without the need of additional ancilla qubits and most importantly making use of non-controlled versions of the generators, which would in principle reduce the number of two-qubit gates and the overall depth of the circuit. It would be also interesting to study the possibility of using variational quantum circuits in order to reproduce a quality approximation of the stabilizer states, or by possibly by encoding the transformations needed to reproduce this stabilizer state via the parameters of the circuit.

Regarding the $U_{NC}U_C$ circuit format in which the Clifford unitary is located on the left, we started by exploring the application of our border procedure in the realm of fully classical simulation of quantum circuits, by leveraging the fact that the Clifford Tableau method is strictly faster than Statevector simulators. To circumvent the fact that a statevector of our stabilizer state would be necessary to proceed with the simulation of the Non-Clifford section of the original circuit, we proposed a composite method based once again on equation \ref{StabProj} and on Stim's capabilities, which reproduces the statevector of the stabilizer state produced by the generators of its stabilizer circuit. In conjunction with the Clifford Tableau in order to acquire the generators, this procedure produces the desired statevector faster than a regular Statevector simulator when the number of qubits ranges between 2 and 9. While the execution time scales poorly as the number of qubits grow, the improvements in execution time for the above mentioned range are significant, specially if the procedure is to be ran multiple times, with executions times of around 5\% of the statevector for 2 and 3 qubits, 7\% for 4 qubits, 11\% for 5 qubits, 16\% for 6 qubits, 28\% for 7 qubits, 47\% for 8 qubits and 71\% for 9 qubits, as illustrated in figure \ref{StimSV_vs_QiskitSV}. While this method is presented in the scope of this work to be used alongside the border detection procedure, it is worth noting how it can be simply used as a general fast statevector simulator for stabilizer circuits under certain conditions, which is a tool that, as far as we know, is not present in Qiskit's implementation of the StabilizerState class. Making use of this alternative statevector simulation method, we can in fact leverage our splitting procedure by efficiently simulating the stabilizer statevector initialization and evolve it through the operator representing the Non-Clifford circuit, which as shown in Figures \ref{CircSim_Stim_v_Qiskit} and \ref{CircSim_Stim_v_Qiskit_100}, represent a significant time improvement in comparison to the traditional statevector simulator used in Qiskit.

Given the results of our proposed composite method for statevector simulation of Clifford+T circuits, we also explored how this would impact execution times of general VQE procedures if the ansatzes fell under this class of circuits. To this end, we introduced several modifications on Qiskit's implementation of the VQE procedure with the intent to include our splitting and simulation procedure as part of the process. The results depicted in Figure \ref{VQE_ExeTimes}, regarding execution times for the $H_2$ molecule eigenstate VQE procedure, seem to be in accordance with the expected speedups identified prior which result from our modification on the statevector simulation, with even an additional technical result revealing that Qiskit's VQE procedure speedup, when configured to run with a statevector simulator, yields significant smaller execution times when set to compute expectation values making use of the \textit{expectation\_value} function of the \textit{Statevector} class.

Some possible improvements regarding these two last major results would probably be entirely focused on sharpening the proposed composite statevector simulation procedure. One such way that we started exploring would be to further improve on the execution time and/or complexity of the algorithm used to reproduce the stabilizer state from the generators, which is a topic that is currently developing with results outpacing Stim's procedure\cite{desilva2024fastalgorithmsclassicalspecifications}. Another possible improvement would be to try and reduce any delays introduced by Qiskit's methods when trying to initialize a statevector simulation, as the \textit{Statevector} class \textit{evolve} method seems to be the one which is less impacted by this effect (when compared to methods such as \textit{initialize} and \textit{set\_statevector} from the \textit{QuantumCircuit} class), but there seems to always be a small time loss associated with such procedures, which contributes to discrepancies with the execution times of certain circuits which should be strictly faster than others (such as subcircuits of a larger circuit).

Finally for circuits in a $U_CU_{NC}$ format, with the Clifford on the right, we briefly went over on how we could significantly reduce the workload to be sent to a quantum backend on typical VQE procedures due to the fact that the expectation value of Pauli observables is conserved under Clifford transformations, and hence the Clifford section of the circuit would instead be used to transform our observables classically, instead of being executed on the quantum backend, with this process having a $O(n^2)$ complexity. Another proposed used case for this scenario also leverages the possibility of replacing the Clifford section by a projector circuit, allowing for the distribution generated by the resultant ancilla qubits to replicate the distribution generated by the full original circuit, with the discussion relating to the effectiveness of this replacement being analogous to the one done in section \ref{Generator Projector Initialization}.

As a final note, we expect that more applications of this Clifford splitting exist, and thus that lays a foundation for possible future works on this matter. One interesting subject that we did not go over during this work but is more or less implicit, would be the possibility of simultaneously leveraging cases of section \ref{Circuits in U_NCU_C format} and \ref{Circuits in U_CU_NC format}, for a circuit in the format $U_CU_{NC}U_C$, which could potentially further increase efficiency on a given scenario, one possible example being the use of statevector simulation coupled with the Clifford transformation of Pauli observables to doubly reduce the size of the effective circuit to be simulated during VQE runs.



\let\oldbibliography\thebibliography
\renewcommand{\thebibliography}[1]{\oldbibliography{#1}
\setlength{\itemsep}{0pt}}
{\footnotesize
\bibliographystyle{mnras}
\bibliography{references.bib}}

\begin{thebibliography}{}
\makeatletter
\relax
\def\mn@urlcharsother{\let\do\@makeother \do\$\do\&\do\#\do\^\do\_\do\%\do\~}
\def\mn@doi{\begingroup\mn@urlcharsother \@ifnextchar [ {\mn@doi@} {\mn@doi@[]}}
\def\mn@doi@[#1]#2{\def\@tempa{#1}\ifx\@tempa\@empty \href {http://dx.doi.org/#2} {doi:#2}\else \href {http://dx.doi.org/#2} {#1}\fi \endgroup}
\def\mn@eprint#1#2{\mn@eprint@#1:#2::\@nil}
\def\mn@eprint@arXiv#1{\href {http://arxiv.org/abs/#1} {{\tt arXiv:#1}}}
\def\mn@eprint@dblp#1{\href {http://dblp.uni-trier.de/rec/bibtex/#1.xml} {dblp:#1}}
\def\mn@eprint@#1:#2:#3:#4\@nil{\def\@tempa {#1}\def\@tempb {#2}\def\@tempc {#3}\ifx \@tempc \@empty \let \@tempc \@tempb \let \@tempb \@tempa \fi \ifx \@tempb \@empty \def\@tempb {arXiv}\fi \@ifundefined {mn@eprint@\@tempb}{\@tempb:\@tempc}{\expandafter \expandafter \csname mn@eprint@\@tempb\endcsname \expandafter{\@tempc}}}

\bibitem[\protect\citeauthoryear{Aaronson \& Gottesman}{Aaronson \& Gottesman}{2004}]{Aaronson_2004}
Aaronson S.,  Gottesman D.,  2004, \mn@doi [Physical Review A] {10.1103/physreva.70.052328}, 70

\bibitem[\protect\citeauthoryear{Backens, Miller-Bakewell, de Felice, Lobski  \& van~de Wetering}{Backens et~al.}{2021}]{Backens_2021}
Backens M.,  Miller-Bakewell H.,  de Felice G.,  Lobski L.,   van~de Wetering J.,  2021, \mn@doi [Quantum] {10.22331/q-2021-03-25-421}, 5, 421

\bibitem[\protect\citeauthoryear{Beverland et~al.,}{Beverland et~al.}{2022}]{beverland2022assessingrequirementsscalepractical}
Beverland M.~E.,  et~al., 2022, Assessing requirements to scale to practical quantum advantage (\mn@eprint {arXiv} {2211.07629}), \url {https://arxiv.org/abs/2211.07629}

\bibitem[\protect\citeauthoryear{Bravyi \& Kitaev}{Bravyi \& Kitaev}{2005}]{Bravyi_2005}
Bravyi S.,  Kitaev A.,  2005, \mn@doi [Physical Review A] {10.1103/physreva.71.022316}, 71

\bibitem[\protect\citeauthoryear{Camps \& Van~Beeumen}{Camps \& Van~Beeumen}{2022}]{9951292}
Camps D.,  Van~Beeumen R.,  2022, in 2022 IEEE International Conference on Quantum Computing and Engineering (QCE). pp 104--113, \mn@doi{10.1109/QCE53715.2022.00029}

\bibitem[\protect\citeauthoryear{Chen, Cotler, Huang  \& Li}{Chen et~al.}{2022}]{chen2022complexitynisq}
Chen S.,  Cotler J.,  Huang H.-Y.,   Li J.,  2022, The Complexity of NISQ (\mn@eprint {arXiv} {2210.07234}), \url {https://arxiv.org/abs/2210.07234}

\bibitem[\protect\citeauthoryear{Duncan, Kissinger, Perdrix  \& van~de Wetering}{Duncan et~al.}{2020}]{Duncan_2020}
Duncan R.,  Kissinger A.,  Perdrix S.,   van~de Wetering J.,  2020, \mn@doi [Quantum] {10.22331/q-2020-06-04-279}, 4, 279

\bibitem[\protect\citeauthoryear{Ekert, Hosgood, Kay  \& Macchiavello}{Ekert et~al.}{}]{qubitguide}
Ekert A.,  Hosgood T.,  Kay A.,   Macchiavello C., , {Introduction to Quantum Information Science}, \url {https://qubit.guide}

\bibitem[\protect\citeauthoryear{Gidney}{Gidney}{2021}]{Gidney_2021}
Gidney C.,  2021, \mn@doi [Quantum] {10.22331/q-2021-07-06-497}, 5, 497

\bibitem[\protect\citeauthoryear{Gottesman}{Gottesman}{1997}]{gottesman1997stabilizercodesquantumerror}
Gottesman D.,  1997, Stabilizer Codes and Quantum Error Correction (\mn@eprint {arXiv} {quant-ph/9705052}), \url {https://arxiv.org/abs/quant-ph/9705052}

\bibitem[\protect\citeauthoryear{Gottesman}{Gottesman}{1998}]{gottesman1998heisenbergrepresentationquantumcomputers}
Gottesman D.,  1998, The Heisenberg Representation of Quantum Computers (\mn@eprint {arXiv} {quant-ph/9807006}), \url {https://arxiv.org/abs/quant-ph/9807006}

\bibitem[\protect\citeauthoryear{Gottesman}{Gottesman}{2009}]{gottesman2009introductionquantumerrorcorrection}
Gottesman D.,  2009, An Introduction to Quantum Error Correction and Fault-Tolerant Quantum Computation (\mn@eprint {arXiv} {0904.2557}), \url {https://arxiv.org/abs/0904.2557}

\bibitem[\protect\citeauthoryear{Kissinger \& van~de Wetering}{Kissinger \& van~de Wetering}{2020}]{Kissinger_2020}
Kissinger A.,  van~de Wetering J.,  2020, \mn@doi [Electronic Proceedings in Theoretical Computer Science] {10.4204/eptcs.318.14}, 318, 229–241

\bibitem[\protect\citeauthoryear{Li, Stein, Krishnamoorthy  \& Ang}{Li et~al.}{2023}]{10.1145/3550488}
Li A.,  Stein S.,  Krishnamoorthy S.,   Ang J.,  2023, \mn@doi [ACM Transactions on Quantum Computing] {10.1145/3550488}, 4

\bibitem[\protect\citeauthoryear{Liu, Gonzales, Huang, Saleem  \& Hovland}{Liu et~al.}{2025}]{liu2025quclearcliffordextractionabsorption}
Liu J.,  Gonzales A.,  Huang B.,  Saleem Z.~H.,   Hovland P.,  2025, QuCLEAR: Clifford Extraction and Absorption for Quantum Circuit Optimization (\mn@eprint {arXiv} {2408.13316}), \url {https://arxiv.org/abs/2408.13316}

\bibitem[\protect\citeauthoryear{Mastel}{Mastel}{2023}]{mastel2023cliffordtheorynqubitclifford}
Mastel K.,  2023, The Clifford theory of the $n$-qubit Clifford group (\mn@eprint {arXiv} {2307.05810}), \url {https://arxiv.org/abs/2307.05810}

\bibitem[\protect\citeauthoryear{Miháliková, Pivoluska, Plesch, Friák, Nagaj  \& Šob}{Miháliková et~al.}{2022}]{Mih_likov__2022}
Miháliková I.,  Pivoluska M.,  Plesch M.,  Friák M.,  Nagaj D.,   Šob M.,  2022, \mn@doi [Nanomaterials] {10.3390/nano12020243}, 12, 243

\bibitem[\protect\citeauthoryear{Quetschlich, Burgholzer  \& Wille}{Quetschlich et~al.}{2023}]{Quetschlich_2023}
Quetschlich N.,  Burgholzer L.,   Wille R.,  2023, \mn@doi [Quantum] {10.22331/q-2023-07-20-1062}, 7, 1062

\bibitem[\protect\citeauthoryear{Sanders, Wallman  \& Sanders}{Sanders et~al.}{2015}]{Sanders_2016}
Sanders Y.~R.,  Wallman J.~J.,   Sanders B.~C.,  2015, \mn@doi [New Journal of Physics] {10.1088/1367-2630/18/1/012002}, 18, 012002

\bibitem[\protect\citeauthoryear{Wang, Fontana, Cerezo, Sharma, Sone, Cincio  \& Coles}{Wang et~al.}{2021}]{Wang_2021}
Wang S.,  Fontana E.,  Cerezo M.,  Sharma K.,  Sone A.,  Cincio L.,   Coles P.~J.,  2021, \mn@doi [Nature Communications] {10.1038/s41467-021-27045-6}, 12

\bibitem[\protect\citeauthoryear{Xu, Benjamin, Sun, Yuan  \& Zhang}{Xu et~al.}{2023}]{xu2023herculeantaskclassicalsimulation}
Xu X.,  Benjamin S.,  Sun J.,  Yuan X.,   Zhang P.,  2023, A Herculean task: Classical simulation of quantum computers (\mn@eprint {arXiv} {2302.08880}), \url {https://arxiv.org/abs/2302.08880}

\bibitem[\protect\citeauthoryear{de Brugière \& Martiel}{de~Brugière \& Martiel}{2024}]{debrugière2024fastershortersynthesishamiltonian}
de Brugière T.~G.,  Martiel S.,  2024, Faster and shorter synthesis of Hamiltonian simulation circuits (\mn@eprint {arXiv} {2404.03280}), \url {https://arxiv.org/abs/2404.03280}

\bibitem[\protect\citeauthoryear{de Silva, Salmon  \& Yin}{de~Silva et~al.}{2024}]{desilva2024fastalgorithmsclassicalspecifications}
de Silva N.,  Salmon W.,   Yin M.,  2024, Fast algorithms for classical specifications of stabiliser states and Clifford gates (\mn@eprint {arXiv} {2311.10357}), \url {https://arxiv.org/abs/2311.10357}

\bibitem[\protect\citeauthoryear{van~de Wetering}{van~de Wetering}{2020}]{vandewetering2020zxcalculusworkingquantumcomputer}
van~de Wetering J.,  2020, ZX-calculus for the working quantum computer scientist (\mn@eprint {arXiv} {2012.13966}), \url {https://arxiv.org/abs/2012.13966}

\makeatother
\end{thebibliography}

\end{document}